\documentclass[aps,twocolumn,amssymb,superscriptaddress,nofootinbib,longbibliography,pra]{revtex4-2}

\usepackage{graphicx}
\usepackage{color}
\usepackage{braket}
\usepackage{verbatim}
\usepackage{amsmath}
\usepackage{mathtools}
\usepackage{braket}
\usepackage{float}
\usepackage{hyperref}
\usepackage{dsfont}
\usepackage{lipsum}
\usepackage{commath}

\begin{document}
\title{The subthreshold issue of fusion-based quantum computing}

\author{Matthias C. L{\"o}bl}
\email{mlo@sparrowquantum.com}
\affiliation{Sparrow Quantum, Nordre Fasanvej 215, DK-2000 Frederiksberg, Denmark}
\author{Love A. M. Pettersson}
\affiliation{Sparrow Quantum, Nordre Fasanvej 215, DK-2000 Frederiksberg, Denmark}
\author{Jan Draga\v{s}evi\'{c}}
\affiliation{Center for Hybrid Quantum Networks (Hy-Q), The Niels Bohr Institute, University~of~Copenhagen,  DK-2100  Copenhagen~{\O}, Denmark}
\author{Susan X. Chen}
\affiliation{Sparrow Quantum, Nordre Fasanvej 215, DK-2000 Frederiksberg, Denmark}
\author{Oliver A. D. Sandberg}
\affiliation{Sparrow Quantum, Nordre Fasanvej 215, DK-2000 Frederiksberg, Denmark}

\begin{abstract}
    Fusion-based quantum architectures are the leading approach to photonic quantum computing.
    However, the sub-threshold regime, where logical error rates must reach the levels required by useful applications, has received little attention.
    We show that in this regime, fusion failure imposes a noise floor on the logical error rate that prevents all-linear-optics architectures from reaching the required rates at low overhead.
    For fusion-based architectures using quantum emitter spins, we show that the noise floor is reduced by orders of magnitude at a lower overhead. 
\end{abstract}
\maketitle

\section{Introduction}
Quantum computers must be built from imperfect physical qubits. Since errors compound over long computations, error correction, where many physical qubits encode a single logical qubit, is required. It is widely believed that most useful quantum computing applications require logical error rates of the order of $\sim10^{-10}$~\cite{Litinski2019, Google2025}, demanding large code distances and making physical qubit count the dominant constraint. 

Photonic quantum computing has emerged as a leading platform under this constraint because photons can be produced in large numbers and naturally provide long-range connections, enabling high-rate quantum low-density parity check (qLDPC) codes~\cite{Panteleev2022, Leverrier2022, Chen2025} which dramatically reduce qubit overhead. Since photon-photon entangling gates are challenging~\cite{Hacker2016}, Bell state measurements (fusions)~\cite{Browne2005, Gimeno2016} have been proposed as probabilistic entangling gates, giving rise to fusion-based quantum computing (FBQC)~\cite{Bartolucci2021}. In FBQC, entangled resource states are generated in parallel and then fused. Fusions drive the computation and their measurement outcomes are combined to realize quantum error correction. 

One key metric in quantum error correction is the error threshold, below which the logical error rate is suppressed exponentially by increasing the code distance~\cite{Kitaev1997, Dennis2002}. In fusion-based architectures, photon loss thresholds have been analysed~\cite{Bartolucci2021, Bombin2023b, Chan2024} and are within reach of state-of-the-art hardware~\cite{psiquantum2025}. However, a threshold does not specify the code distance needed to reach a given logical error rate. 
The overhead required to reach a target error rate has been studied in circuit-based architectures~\cite{Bravyi2013, Mayer2025, Google2025, Benhemou2025, Gidney2023}, but the FBQC analogue has received little attention. 

In FBQC, fusion failure and photon loss both produce erasures, and are typically combined into a single erasure budget, obscuring the fact that fusion failure, unlike loss, does not improve with hardware engineering. Fusion failure thus imposes a noise floor that persists even at zero physical loss — an obstacle previously not emphasised in the FBQC literature~\cite{Bartolucci2021, Bombin2023b, Chan2024, Paesani2022}. We show that this floor is a major constraint in the sub-threshold regime. The fusion failure rate can be reduced via boosting with ancillary photons, but at exponential photon cost~\cite{Grice2011}, eventually destroying loss tolerance~\cite{Bartolucci2021}. 
In practical terms, FBQC at a fixed code distance may not reach the required logical error rates at any physical loss rate.

We analyse the sub-threshold regime of several fusion-based architectures numerically, including all-linear-optics architectures~\cite{Bartolucci2021, Bombin2023b} and recently proposed architectures based on quantum emitters~\cite{Chan2024, Chan2025, Hilaire2024}. In the latter architectures, quantum emitters generate spin-photon entanglement deterministically~\cite{Lindner2009}, which is converted into spin-spin entanglement by repeating linear optics fusions until success~\cite{Chan2024}. We find that the quantum-emitter-based architectures have a strongly suppressed noise floor sub-threshold. Compared to all-linear-optics architectures~\cite{Bartolucci2021, Bombin2023b} the overhead for reaching $10^{-10}$ logical error rate is reduced by orders of magnitude.
Quantum emitters thus open a route to useful FBQC at realistic resource overheads.

\section{Architectures}
\begin{figure*}
    \centering
    \includegraphics[width=1.0\textwidth]{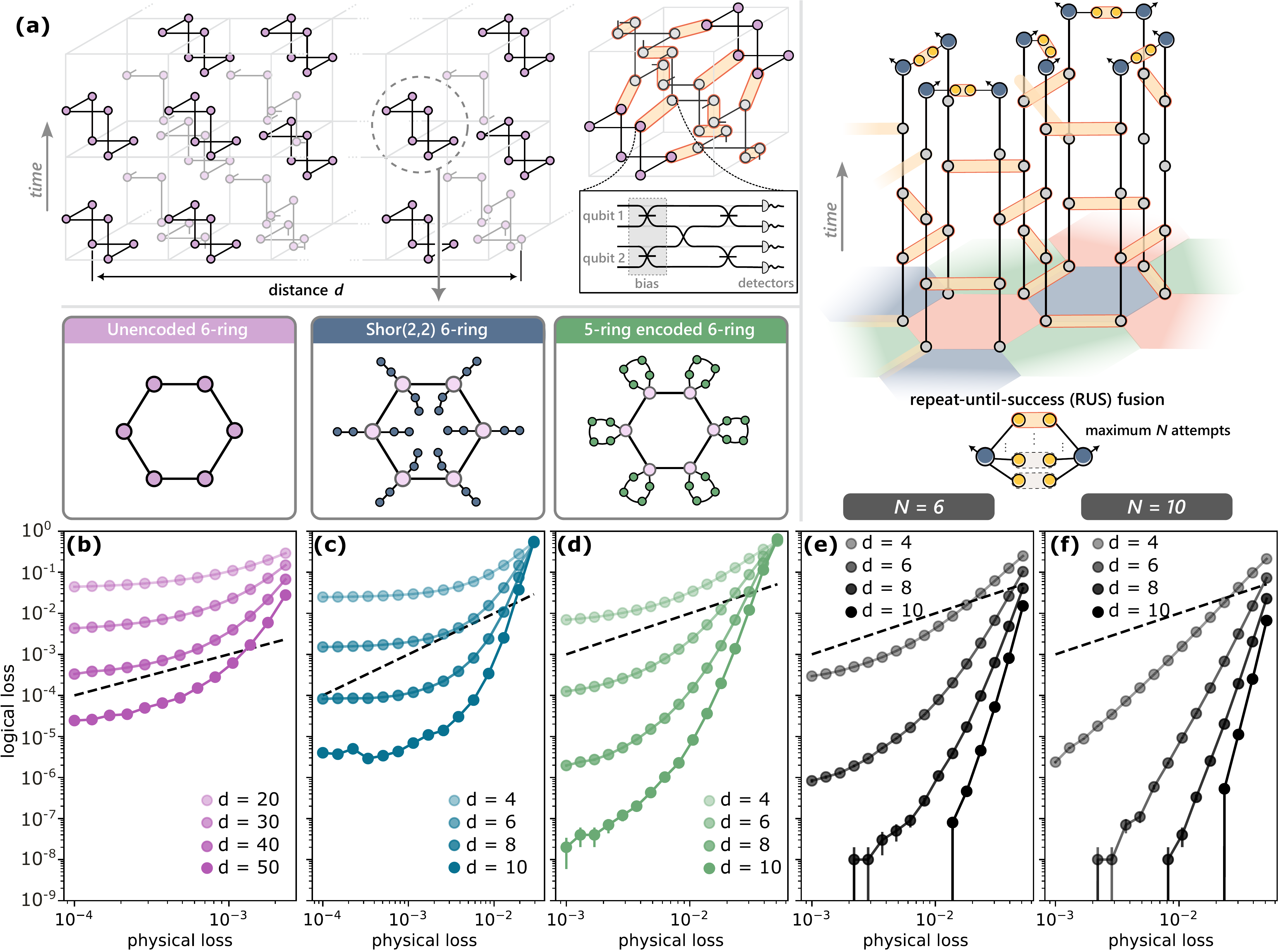}
    \caption{\textbf{(a)} Illustration of the $6$-ring fusion-based architecture~\cite{Bartolucci2021} (left) and the FFCC RUS architecture~\cite{Chan2024} (right). Photons are illustrated as circles and quantum emitter spins are illustrated as circles with an arrow. Type-II fusions are illustrated by the orange ovals with an exemplary setup shown for dual-rail encoded qubits. In \textbf{(b-f)} we show the logical loss rate as a function of the physical loss rate (subthreshold regime), with the dashed black lines indicating where the physical and logical loss rates are equal. The simulations are shown for (b) the six-ring architecture with static bias arrangement and singly-boosted fusion~\cite{Bombin2021}, (c) the Shor(2,2)-encoded six-ring architecture~\cite{Bartolucci2021} with singly boosted fusion and random code orientation, (d) the five-ring-encoded~\cite{Bell2022} six-ring architecture with unboosted fusion and static code orientation, (e, f) the foliated Floquet color code architecture with repeat-until-success fusions~\cite{Chan2024, Chan2025}, where $N$ is the maximum number of repeated fusion attempts.}
    \label{fig:sub}
\end{figure*}
We consider different fusion-based architectures that are illustrated in Fig.~\ref{fig:sub}(a). First, we consider architectures where entangled photonic resource states are arranged and fused in a fault-tolerant fusion network~\cite{Bartolucci2021, Bombin2023b} with some spatial extension (distance), $d$. We call these architectures \textit{all-linear optic} because the resource states can be generated probabilistically from single photons and linear optics circuits with measurements~\cite{Li2015}. The fusions are realized by linear optics Bell state measurements that consume the two measured photons (type-II fusions~\cite{Browne2005, Gimeno2016}). Upon success, they measure the Bell state stabilizer generators $\langle XX,ZZ\rangle$. Upon fusion failure, the fusions measure both photons in either the Pauli $X$ or the Pauli $Z$ basis, depending on the fusion \textit{bias}. In this way, one of the two stabilizer generators is obtained as a product of the two single-qubit measurements, and the other one is erased. If one or more photons are lost, both stabilizer generators are erased. Erasure due to fusion failure can be reduced by boosting the fusion~\cite{Grice2011}, where $2^n-2$ ancillary photons~\footnote{The fusion success probability could also be boosted using quantum-emitter-based nonlinearities, a method that, however, is experimentally less established~\cite{Nielsen2026}.} are required to obtain a fusion failure probability of $p_f=1/2^n$. If the fusion bias is chosen randomly (random bias arrangement), the erasure probability of every stabilizer generator is therefore~\cite{Bartolucci2021}
\begin{equation}
    \label{eq_ers}
    p_{ers} = 1-(1-p_f/2)\cdot\eta^{1/p_f},
\end{equation}
where $\eta$ is the efficiency (probability that a photon is not lost). Alternatively, the fusion failure basis can be fixed (static bias arrangement) or it can be adapted based on previous fusion outcomes (dynamic bias arrangement), which can improve performance~\cite{Bombin2023b}. In this work, we consider the $6$-ring architecture from Ref.~\cite{Bartolucci2021} with random or static bias arrangement~\cite{Bombin2023b}, with Shor(2,2) encoding~\cite{Bartolucci2021}, and with an optimized $5$-ring encoding of the same size~\cite{Pettersson2025, Bell2022}. In the last case, the measurement basis is adapted conditioned on measurement outcomes within the same encoded fusion. The other architectures are non-adaptive.

Furthermore, we consider architectures that use quantum emitter spins as data qubits and repeat-until-success (RUS) fusions to realize entangling gates between the spins~\cite{Chan2024, Chan2025}. The particular architecture we consider is based on the foliated Floquet color code~\cite{Paesani2022, Bombin2023c}, and we call it \textit{FFCC RUS}. In this architecture, spin-entangled photons are generated on demand~\cite{Lindner2009} and fused with an unboosted fusion, resulting in entangling gates between the spins upon fusion success. If a fusion fails, new spin-entangled photons are generated and the fusion is repeated until success, with some maximum number $N$ of allowed physical fusion attempts. The strength of RUS architectures is that the probability of fusion failure is reduced to $p_f=1/2^N$ (in the absence of photon loss~\footnote{We refer the reader to Eq. B5 in Ref~\cite{Chan2024} for detailed erasure probabilities when loss is present.}) without exponentially many ancillary photons as in boosted fusion~\cite{Grice2011}. In fact, the average number of fusion photons is given by
\begin{equation}
\label{eq_n_rus}
    N_{RUS} = 2\sum_{k=1}^{N}k/2^k,
\end{equation}
converging to a value of just $4$ for $N\rightarrow\infty$. Furthermore, repeating fusion upon demand is only minimally adaptive. In contrast to more adaptive strategies such as dynamic bias arrangement~\cite{Bombin2023b} or Macromux~\cite{Birchall2026}, the adaptiveness is also local, making parallelization particularly simple.

\section{Numerical Simulations}
\begin{figure*}
    \centering
    \includegraphics[width=1.0\linewidth]{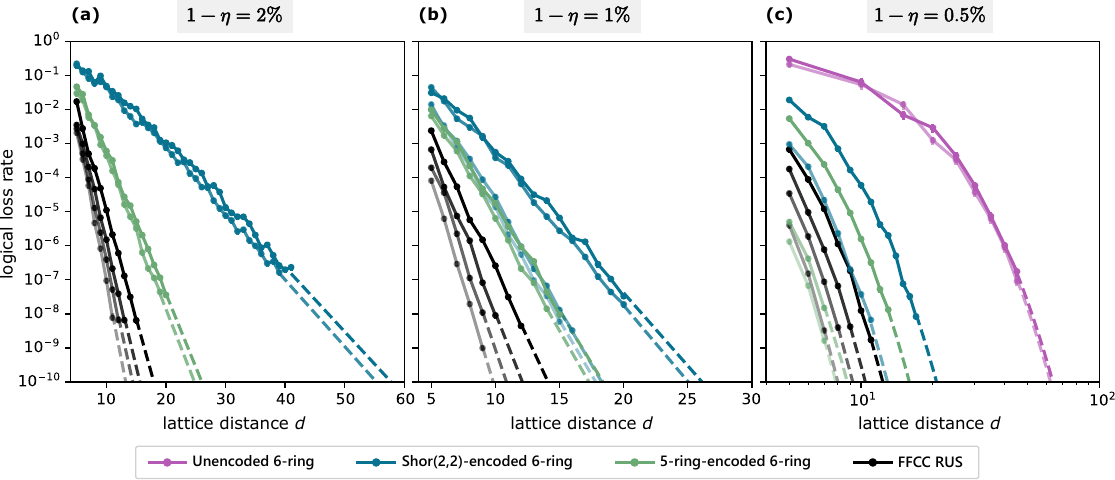}
    \caption{Logical loss rate as a function of fusion network size, $d$. \textbf{(a)} Simulation for $2\,\%$ physical loss. The FFCC RUS architecture~\cite{Chan2024, Chan2025} is plotted for $N=5$ (solid), $N=6,7,\infty$ (most transparent). The Shor(2,2)-encoded and the $5$-ring-encoded~\cite{Pettersson2025} $6$-ring architectures~\cite{Bartolucci2021, Bombin2023b} are plotted for random and static (more transparent) code orientation, where we have used unboosted fusion for the $5$-ring encoding and singly boosted fusion~\cite{Grice2011} for the Shor(2,2) encoding. The FFCC RUS architecture has a favorable scaling in comparison to both encoded $6$-ring architectures.\footnote{The error bars all have about the same size on the logarithmic scale because we sample until the logical loss rate has a relative uncertainty of $0.2$. This way, we dynamically adapt the number of samples and only use as many as needed.} \textbf{(b)} Simulation for $1\,\%$ physical loss. For the encoded $6$-ring architecture, we also consider doubly boosted fusion (more transparent). \textbf{(c)} Simulation for $0.5\,\%$ physical loss, where the unencoded $6$-ring scheme~\cite{Bartolucci2021, Bombin2023b} with doubly boosted fusion is below its threshold as well (static bias arrangement is more transparent). For better visibility, we only show the simulation for random code orientation for the encoded $6$-ring architectures (higher boosting is more transparent).}
    \label{fig:sub_scaling}
\end{figure*}

We simulate the logical loss rate as a function of the physical loss rate by repeatedly sampling erasures due to photon loss and fusion failure. When a fusion measurement outcome is erased, the corresponding parity-check cells must be merged into a super-cell~\cite{Stace2009}. Erasure can be simulated as a percolation problem on the syndrome graph~\cite{Stace2009, Bombin2023b}. Each erased $XX$ or $ZZ$ fusion outcome corresponds to a syndrome graph edge, and when a connected cluster of erasures spans (percolates) the fusion network, there is a logical loss. For random bias arrangement~\cite{Bartolucci2021}, $XX$ and $ZZ$ outcomes have the same probability of erasure (Eq.~\eqref{eq_ers}), and logical loss can be simulated by standard bond percolation~\footnote{Knowing the syndrome graph percolation thresholds, one can, for instance, reproduce Fig. 4 from Ref.~\cite{Bartolucci2021} with a few lines of code that we attach.}. In the case of the FFCC RUS architecture~\cite{Chan2024, Chan2025} and the architecture with static bias arrangement~\cite{Bombin2023b} $XX$ and $ZZ$ outcomes have different erasure probabilities, leading to inhomogeneous bond percolation. Both types of bond percolation simulations~\footnote{We use the source code from~\cite{Lobl2023github, Lobl2024b} to which we added more features for non-uniform bond percolation.} have linear time complexity~\cite{NewmanZiff2001} and thus are not computationally demanding.

As a consistency check, we first determine loss thresholds of the different architectures by identifying the point where logical loss rates intersect for different lattice sizes. The thresholds agree with the values reported in the literature~\cite{Bartolucci2021, Chan2024}, except for the unencoded $6$-ring architecture with static bias arrangement~\cite{Bombin2023b}, where our threshold is about $0.1\,\%$ lower. We find thresholds of $7.0\%, 7.2\%, 7.3\%, 7.3\%$ for the FFCC RUS architecture~\cite{Chan2024} with $N=6, N=7, N=10, N\rightarrow\infty$; $2.7\%$ for the Shor(2,2)-encoded $6$-ring architecture from~\cite{Bartolucci2021} for random or static code orientation and singly boosted fusion; $0.38\%$ $(0.82\%)$ for the unencoded $6$-ring architecture with static bias arrangement and singly (doubly) boosted fusion~\cite{Bombin2023b}; and $0.78\%$ for the unencoded $6$-ring architecture with random bias arrangement and doubly boosted fusion~\cite{Bartolucci2021}~\footnote{For the FFCC RUS architecture, we used the model described in Appendix B of Ref.~\cite{Chan2024}. Other models exploiting quantum emitter reinitialization (Appendix C of Ref.~\cite{Chan2024}) would require a higher level of real-time adaptivity.}. For threshold values for other physical noise sources, we refer the reader to Refs.~\cite{Bartolucci2021, Bombin2023b, Chan2025}.

We investigate the subthreshold regime for the different architectures as shown in Fig.~\ref{fig:sub}(b-f). For the unencoded $6$-ring architecture with static bias arrangement~\cite{Bombin2023b} in Fig.~\ref{fig:sub}(b) and the Shor(2,2)-encoded $6$-ring achitecture~\cite{Bartolucci2021} in Fig.~\ref{fig:sub}(c), we use singly boosted fusion~\cite{Grice2011} because there is no loss threshold otherwise~\cite{Bartolucci2021}. In both cases, the logical loss rate bends and plateaus towards zero physical loss. The plateau is due to the finite fusion failure probability, although it is just $p_f=1/4$ due to boosting~\cite{Grice2011}. Due to the constant rate of fusion failure, there is a finite probability, for any fixed size $d$, that a chain of erasures percolates the syndrome graph. This results in a finite logical loss rate (plateau), regardless of how low the physical loss is. As a result, a fusion network with distance $d>40$ is required to even reach the break-even point where logical loss equals physical loss (see Fig.~\ref{fig:sub}(b)). Furthermore, in Fig.~\ref{fig:sub}(c), one can see that the break-even point is rather a break-even regime: for every fixed size $d$, there is a physical loss rate below which the logical loss rate again exceeds the physical loss rate. An arbitrarily low logical loss rate can still be reached by increasing $d$, but the required values for $d$ can become very large.

The described issue in the subthreshold regime is also present for the two locally adaptive architectures in Fig.~\ref{fig:sub}(d,e,f). However, it is much weaker and much lower logical error rates can be reached at smaller values of $d$. For the FFCC RUS architecture in Fig.~\ref{fig:sub}(e,f), this improvement can be intuitively explained: at zero loss, an RUS fusion suffers erasure due to fusion failure only if all $N$ physical fusion attempts fail. Increasing $N$, thus exponentially reduces the probability $1/2^N$ that RUS fusion fails, strongly reducing the plateauing of the logical error rate towards zero loss. In contrast to boosting with ancillary photons~\cite{Grice2011}, this reduction of fusion failure is achieved without an exponential number of ancillary photons, maintaining high tolerance to photon loss.

To obtain a quantitative comparison between the different architectures, we fix the physical loss rate $1-\eta$ to $0.5\%, 1\%, 2\%$ and estimate the value of $d$ where the logical loss rate drops below $10^{-10}$~\cite{Google2025}. For all-linear optics architectures, we choose the boosting~\cite{Grice2011} level for which the overall probability of erasure in Eq.~\eqref{eq_ers} is minimized. We find that singly boosted fusion ($p_f=1/4$) is best for $1-\eta=2\%$, doubly boosted fusion ($p_f=1/8$) is best for $1-\eta=0.5\%$ and $1-\eta=1\%$. We also simulate singly-boosted fusion in the two cases where doubly-boosted fusion leads to a lower erasure rate, as the overhead may be lower when using fewer ancillary photons. The corresponding simulations are shown in Fig.~\ref{fig:sub_scaling}. To reach a logical loss rate of $10^{-10}$, the FFCC RUS architecture~\cite{Chan2024, Chan2025} requires the smallest distance $d$ of the fusion network, followed by the $5$-ring-encoded $6$-ring architecture with unboosted fusion and adaptive basis choice~\cite{Pettersson2025}. The Shor(2,2)-encoded (unencoded) $6$-ring architecture~\cite{Bartolucci2021, Bombin2023b} has an overhead that is greater by about one (two) orders of magnitude when counting the photons of the $d^3$ unit cells of the fusion network. This comparison only takes into account the size of the fusion network in space and time, completely ignoring the additional overhead of generating photonic resource states ($6$-ring, Shor(2,2)-encoded $6$-ring, $5$-ring-encoded $6$-ring) with linear optics~\cite{Lee2023}.

To obtain a more physical comparison, we convert $d$ to the total number of photons, including those consumed in generating the resource states. In the case of FFCC RUS, we use Eq.~\ref{eq_n_rus} to compute the overhead as $R=d^3\times 6N_{RUS}$, where the factor of six is half the degree of the syndrome graph, avoiding double counting. In the linear optics case, the overhead is $R=d^3\times \left(N_{r} + N_f\right)$, where $N_r$ is the average number of photons required to generate the resource states within one unit cell, and $N_f$ is the per unit cell number of ancillary photons for boosting~\cite{Grice2011}. We assume the same level of boosting in resource state generation and the fault-tolerant fusion network. Furthermore, we only count two photons for every Bell pair used in boosted fusion~\cite{Grice2011}, not counting the additional overhead when generating these Bell states by linear optics~\cite{Fldzhyan2021}.

In the all-linear optics case, we consider a scheme where resource states are built from three-qubit GHZ states in a divide-and-conquer fashion~\cite{Lee2023}, and the three-qubit GHZ states are generated from single photons via linear optics and post-selection~\cite{Li2015}~\footnote{One also could consider the architectures from Refs.~\cite{Bartolucci2021, Bombin2023b} using quantum emitters for resource state generation~\cite{Li2022, Lobl2024, Wein2024, Manohar2026}. However, accumulation of quantum emitter noise during the resource-state generation is likely quite problematic in these schemes. Furthermore, several interacting quantum emitters are often required to generate useful graph states~\cite{Li2022, Manohar2026}.}. We obtain $N_r = n_{GHZ}\cdot f_{GHZ} + n_b$, where $n_{GHZ} (n_b)$ is the average number of three-qubit GHZ states (ancilla boosting photons) used to build the resource state. Furthermore, $f_{GHZ}=6\cdot32$ represents the average number of photons consumed to generate a three-qubit GHZ state with the standard GHZ-state factory~\cite{Li2015} using six input photons and having $1/32$ success rate. To estimate $n_{GHZ}$ and $n_b$, we use the source code from Ref.~\cite{Lee2023}. Here, the effect of fusion failure on the probability of successful resource state generation is considered, ignoring heralded loss from missing detection events in the resource state generation. Noise accumulation in the all-linear-optics resource state generation is also not taken into account. The considered loss rates thus refer to the final resource state. Both are optimistic assumptions for the all-linear approach.

\begin{figure}
    \centering
    \includegraphics[width=1.0\linewidth]{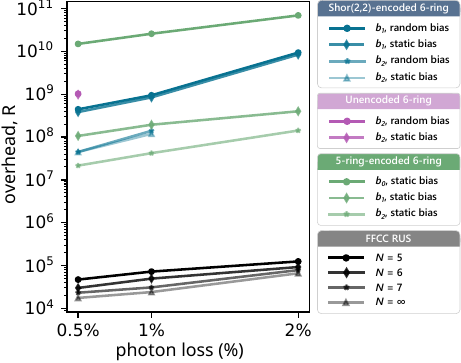}
    \caption{Overhead $R$ in terms of the number of consumed photons to reach a logical loss rate of $10^{-10}$. The overhead is shown for three photon loss rates and the different architectures (FFCC RUS~\cite{Chan2024, Chan2025}, Shor(2,2)-encoded $6$-ring fusion network~\cite{Bartolucci2021,Bombin2023b}, unencoded $6$-ring fusion network~\cite{Bartolucci2021, Bombin2023b}, and $5$-ring-encoded $6$-ring fusion network~\cite{Pettersson2025, Bell2022}). Unboosted, singly and doubly boosted fusion are indicated as $b_0,b_1,b_2$. Bias refers to static (random) bias arrangement for the unencoded $6$-ring architecture~\cite{Bombin2023b} or static (random) code orientation for the encoded architectures~\cite{Bartolucci2021}.}
    \label{fig:overhead}
\end{figure}

We summarize the overhead $R$ in Fig.~\ref{fig:overhead}. The value of $d$, where the logical loss rate crosses $10^{-10}$, is rounded to the next integer for this estimation. For reaching $10^{-10}$, we find that the FFCC RUS architecture~\cite{Chan2024, Chan2025} has a significantly lower overhead than the considered all-linear-optics architectures~\cite{Bartolucci2021, Bombin2023b}. The difference in overhead is several orders of magnitude across various photon-loss rates. This is even the case for the optimized (adaptive) $5$-ring-encoded $6$-ring architecture~\cite{Pettersson2025}, where the required value for $d$ is reduced, but resource state generation dominates the overhead.

\section{Summary and Outlook} For different fusion-based quantum computing architectures, we benchmarked the subthreshold regime in the presence of photon loss and fusion failure. We identified a noise floor caused by fusion failure and we estimated what is required to reach a logical loss rate of $10^{-10}$. We find that non-adaptive all-linear-optics architectures~\cite{Bartolucci2021, Bombin2023b} require a fusion network of very large distance, resulting in significant overhead. The distance can be reduced by using encoded resource states~\cite{Bell2022, Pettersson2025}, which, however, shifts the overhead to resource-state generation~\cite{Lee2023}. Reducing the photon overhead in all-linear-optics quantum computing will therefore likely require adding a high degree of adaptivity on the level of fault-tolerance (dynamic bias arrangement~\cite{Bombin2023b} or Macromux~\cite{Birchall2026}) or on the level of resource state generation~\cite{Staudacher2026, Duan2005}. In both cases, this adds an overhead to the classical control.

In contrast, the FFCC RUS architecture~\cite{Chan2024, Chan2025} does not require the generation of large resource states, as it uses quantum emitter spins as data qubits. FFCC RUS reduces the fusion-failure-induced noise floor by repeating fusions on demand, keeping photon-loss-induced erasures to a minimum. As a result, much less overhead is required to achieve low logical loss rates, making such emitter-based architectures~\cite{Chan2024, Chan2025, Hilaire2024} very promising candidates for fusion-based quantum computing. We emphasize that emitter-based architectures are not limited to the Floquet color code, but a large class of quantum low-density-parity-check (qLDPC) codes can be implemented similarly~\cite{Chen2025}. As qLDPC codes can encode a higher rate of logical qubits per physical qubits~\cite{Panteleev2022, Leverrier2022}, they may be a route to further reduce overhead.

\section*{Acknowledgements}
We thank Anders S. S{\o}rensen, Peter Lodahl, and Stefano Paesani for fruitful discussions. We thank Amazon Web Services for free computing time for this project.

\bibliography{main.bbl}

\end{document}